\documentclass[9pt,twocolumn,twoside]{opticajnl}
\journal{opticajournal} 

\setboolean{shortarticle}{true}

\usepackage{subcaption}

\title{Bifurcation structure of soliton self-injection locking in microresonators}

\author[1]{S. Deshmukh}
\author[1]{ T. M. Schneider}
\author[2,*]{A. Tikan}

\affil[1]{Emergent Complexity in Physical Systems Laboratory (ECPS), Swiss Federal Institute of Technology Lausanne (EPFL), CH-1015 Lausanne, Switzerland}
\affil[2]{Laboratoire Temps-Fréquence, Université de Neuchâtel, Avenue de Bellevaux 51, Neuchâtel, Switzerland}

\affil[*]{alexey.tikan@unine.ch}

\begin{abstract} 
Self-injection locking (SIL) of a diode laser to a high quality-factor microresonator has recently become increasingly important in hybrid integrated photonics, providing access to compact sub-Hz linewidth lasers. It was also shown to facilitate the access to dissipative Kerr solitons - the key to a low-noise coherent frequency comb on a photonic chip. However, the existence and stability ranges of SIL soliton states in experimentally controlled parameters are still not fully understood. Here we study the bifurcation structure of solutions in a model of soliton SIL in the weak-backscattering limit. We show that SIL produces soliton-number-dependent existence ranges of multi-soliton solutions in free-laser detuning and feedback phase parameters. We identify exclusive single-soliton existence regions and demonstrate dynamical access to single solitons in this region by direct numerical simulations using prescribed parameter sweeps.
\end{abstract}

\setboolean{displaycopyright}{false}

\begin{document}
\maketitle
Microresonator-based optical frequency combs (also known as microcombs) have emerged as a compact route to broadband, phase-coherent spectra by exploiting Kerr nonlinearity in high-$Q$ cavities~\cite{herr2026FrequencyCombsCoherent}. In the anomalous-dispersion regime, dissipative Kerr solitons (hereafter referred to as soliton) provide a particularly important operating state, characterized by stable ultrashort pulses with low-noise comb lines and well-defined repetition rates~\cite{kippenbergDissipativeKerrSolitons2018,herr2014TemporalSolitonsOptical}. This has enabled a rapidly expanding set of applications spanning timing, microwave photonics, sensing, spectroscopy, and communications~\cite{sun2023ApplicationsOpticalMicrocombs}. A coherent frequency comb can be generated with any number of solitons circulating in the cavity; however, for most applications the single-soliton regime is favored because it produces a smooth $\mathrm{sech}^2$-shaped spectral envelope. In contrast, multi-soliton states yield a spectrum with pronounced intensity fluctuations between neighboring comb lines~\cite{Bao:17}. However, despite some progress~\cite{yu2021SpontaneousPulseFormation,ulanov2024SyntheticReflectionSelfinjectionlocked}, in practice, reliably hitting a target soliton number during a laser detuning scan can be difficult, whether due to advanced design and fabrication or experimental complexity.

Self-injection locking (SIL) offers a compact route to address this challenge by using the microresonator as a frequency-selective feedback element for a laser diode. SIL is realized by allowing backward-propagating light from the microresonator to re-enter the laser cavity, often by omitting the optical isolator. The backward field can arise intrinsically from Rayleigh backscattering in the resonator~\cite{gorodetsky2000RayleighScatteringHighQ}, or be engineered through synthetic reflection in corrugated resonators~\cite{ulanov2024SyntheticReflectionSelfinjectionlocked}. Resonant feedback is known to narrow and stabilize semiconductor lasers~\cite{lang1980ExternalOpticalFeedback,kazarinov1987RelationLineNarrowing,dahmani1987FrequencyStabilizationSemiconductor,laurent1989FrequencyNoiseAnalysis,kondratiev2017SelfinjectionLockingLaser}. More recently, in microresonator systems, the same feedback mechanism has enabled direct diode pumping of soliton microcombs~\cite{pavlov2018NarrowlinewidthLasingSolitona}. Together with rapid progress in chip-scale and microresonator-stabilized lasers~\cite{jin2021HertzlinewidthSemiconductorLasersa,liang2010WhisperinggallerymoderesonatorbasedUltranarrowLinewidth,corato-zanarella2023WidelyTunableNarrowlinewidth,snigirev2023UltrafastTunableLasers,siddharth2022UltravioletPhotonicIntegrated}, these advances have motivated a growing body of theoretical and experimental studies of the underlying soliton dynamics~\cite{kondratiev2023RecentAdvancesLasera}.

\begin{figure*}[!t]
\centering\includegraphics[width=0.8\textwidth]{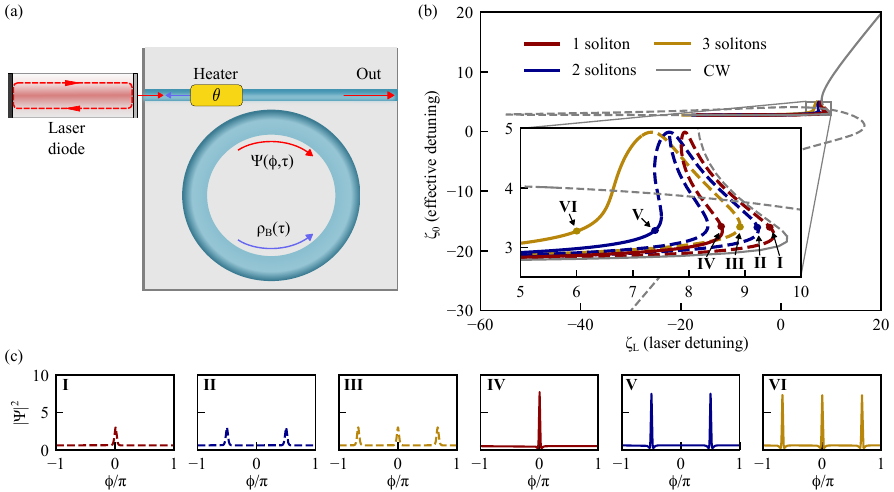}
\caption{\textbf{SIL soliton branches overlaid with the nonlinear tuning curve.} 
(a) Schematic of a diode laser self-injection locked to a high-$Q$ microresonator through weak back-reflection. 
(b) Bifurcation diagram of one-, two-, and three-soliton branches in the $(\zeta_{\mathrm{L}},\zeta_0)$ plane, overlaid on the CW tuning curve. The nearly horizontal segment of the tuning curve corresponds to the locking region, where the effective detuning $\zeta_0$ remains almost constant over a broad range of free-laser detuning $\zeta_{\mathrm{L}}$. Solid and dashed curves denote stable and unstable equilibria, respectively. The inset highlights the detuning interval $\zeta_{\mathrm{L}}\in[7.51,8.60]$ in which the single-soliton branch is the only stable soliton branch. 
(c) Representative intracavity intensity profiles of the forward field $|\Psi(\phi)|^2$ at the points marked in (b) (I--III unstable and IV--VI stable); these profiles coincide with the single-LLE soliton solutions at the corresponding effective detuning $\zeta_0$. Parameters are $d_2=0.01$, $\theta=0.6\pi$, $K=100$, $f^2=4$, and $\beta=0.01$.}
\label{fig_1}
\end{figure*} 

In conventional laser-driven microresonator experiments, detuning scans typically display in sequence CW background, Turing patterns, spatiotemporal chaos, and breathers before reaching the soliton regime, often yielding multi-soliton states with scan-to-scan variation of the resulting soliton number~\cite{herr2014TemporalSolitonsOptical}. The availability of multi-soliton states is explained by a bifurcation analysis of the Lugiato-Lefever equation (LLE) model of the microresonator, whose $N$-soliton solution branches coexist over a common detuning range, organized within a foliated snaking bifurcation structure~\cite{Parra-Rivas2018}. By contrast, SIL detuning scan experiments often show a direct CW-to-soliton transition without an intermediate chaotic stage and allow prescribed $N$-soliton states, including a single soliton, to be accessed for a suitable feedback coupling and phase~\cite{voloshin2021DynamicsSolitonSelfinjection}. This contrast calls for a systematic bifurcation analysis of the coupled laser-microresonator system.

In this letter, we present a bifurcation analysis of the localized equilibrium (dissipative Kerr) soliton solutions of the coupled LLE--diode-laser model in the small-backscattering limit. We discover a slanted snaking bifurcation structure of soliton solution branches, so that the existence range of $N$-soliton solutions in the free-laser detuning and the feedback phase parameters depends on the soliton number $N$. We characterize the stability of these $N$-soliton solutions and identify regions in parameter space where the single soliton is the only stable soliton attractor. We then demonstrate dynamic access to single solitons through direct numerical simulations with prescribed sweeps of the laser detuning and feedback phase.

\textbf{Model. }
The system consists of a laser diode butt-coupled to a high-$Q$ microresonator on a photonic chip, with no intervening optical isolator (Fig.~\ref{fig_1}(a)). The forward-propagating intracavity field $\Psi$ gives rise, via weak back-reflection, to a counter-propagating field $\rho_{\mathrm{B}}$, which in turn returns to the laser cavity and couples with the laser field, closing the self-injection feedback loop. The principal control parameters are the laser pump frequency, the pump power, and the round-trip feedback phase $\theta$, which sets how the returned light interferes with the intracavity laser field and is typically tuned with an integrated heater.

We describe the system by the following normalized coupled equations~\cite{wang2022SelfregulatingSolitonSwitchinga}:
\begin{equation}
\begin{aligned}
\frac{\partial \Psi}{\partial \tau} &= -(1 + i\zeta_0)\Psi + i\frac{d_2}{2} \frac{\partial^2 \Psi}{\partial \phi^2} + i|\Psi|^2\Psi + f, \\
\frac{d\rho_{\mathrm{B}}}{d\tau} &= -(1 + i\zeta_0 - 2iP)\rho_{\mathrm{B}} + i\beta\rho, \\
\zeta_0 &= \zeta_{\mathrm{L}} + K \mathrm{Im} \left[ e^{i\theta} \frac{\rho_{\mathrm{B}}}{i\beta f} \right],
\label{eq1:sil_lle}
\end{aligned}
\end{equation}
where $\phi\in[-\pi,\pi]$ is the azimuthal angle around the resonator and $\tau=t/\tau_{\mathrm{ph}}$ is the slow time in units of the photon lifetime $\tau_{\mathrm{ph}}=2/\kappa$ ($\kappa$ is the total loss rate). $\Psi(\phi,\tau)$ is the slowly varying envelope of the forward field, $\rho=(2\pi)^{-1}\int_{-\pi}^{\pi}\Psi\,d\phi$ its cavity average, and $P=(2\pi)^{-1}\int_{-\pi}^{\pi}|\Psi|^2\,d\phi$ its mean power; $\rho_{\mathrm{B}}$ is the corresponding average of the counter-propagating (backward) field envelope. The free-running laser detuning $\zeta_{\mathrm{L}}=2(\omega_0-\omega_{\mathrm{L}})/\kappa$ measures the laser's unperturbed emission frequency $\omega_{\mathrm{L}}$ relative to the cold-cavity resonance $\omega_0$, in units of the half-linewidth $\kappa/2$; it is the control parameter scanned in detuning sweeps. The effective detuning $\zeta_0=2(\omega_0-\omega_{\mathrm{p}})/\kappa$ is defined likewise, but with the actual frequency $\omega_{\mathrm{p}}$ at which the locked laser pumps the resonator; it is the detuning experienced by the microresonator. The remaining parameters are the normalized second-order dispersion $d_2=2D_2/\kappa$ ($D_2$ the integrated second-order dispersion coefficient), the pump amplitude $f$ (so that $f^2$ is the normalized pump power), the effective feedback gain $K$, the feedback phase $\theta$, and the normalized backscattering $\beta=2\overline{g}_{\mathrm{L}}/\kappa$, where $\overline{g}_{\mathrm{L}}$ is the coupling strength between the zeroth modes of the forward and backward fields. 

The system Eqs.~\ref{eq1:sil_lle} is a reduction, under two key approximations, of the coupled LLE and diode-laser equations describing the SIL system~\cite{wang2022SelfregulatingSolitonSwitchinga}. First, the backscattering is assumed to be weak ($\beta\ll1$), so the higher azimuthal modes of the backward field decouple and only its cavity average $\rho_{\mathrm{B}}$ survives, obeying the ordinary differential equation above. Second, the laser intensity relaxes much faster than the intracavity field and is adiabatically eliminated, leaving the phase-locking condition that pins the effective detuning $\zeta_0$ to the free-laser detuning $\zeta_{\mathrm{L}}$ plus a feedback term. The feedback gain $K$, which sets the locking bandwidth, is assumed to be large because the laser-cavity $Q$ lies well below the microresonator's. Consequently the forward field obeys the standard LLE, with the feedback entering only through the effective detuning $\zeta_0$ it selects; the forward-field component of every SIL soliton therefore coincides with a single-LLE soliton at the same detuning $\zeta_0$ and pump $f$.

\begin{figure}[htbp]
\centering\includegraphics[width=0.8\columnwidth]{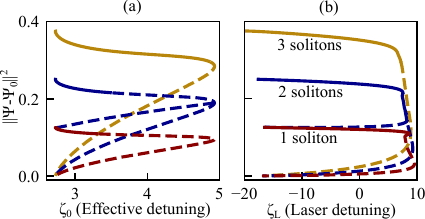}
\caption{\textbf{Soliton branches in effective and free-laser detuning.} Bifurcation diagram of the soliton solution branches in (a) effective detuning $\zeta_0$ and (b) free-laser detuning $\zeta_{\mathrm{L}}$, using the CW background-subtracted norm $\|\Psi-\Psi_0\|^2$. The forward intracavity field $\Psi(\phi)$ of the stationary soliton solution satisfies the single LLE, revealing the familiar foliated snaking bifurcation structure of the LLE solitons in effective detuning $\zeta_0$ (with modified stability features due to the feedback coupling induced instabilities); applying the SIL locking condition maps the same equilibria to free-laser detuning $\zeta_{\mathrm{L}}$, displaying soliton-number-dependent existence ranges in $\zeta_{\mathrm{L}}$. Parameters are $d_2=0.01$, $\theta=0.6\pi$, $K=100$, $f^2=4$, and $\beta=0.01$.}
\label{fig_2}
\end{figure}

\textbf{Soliton branches on the tuning curve.}
During a typical laser detuning sweep of the SIL system, the effective detuning $\zeta_0$ tracks the free-laser detuning $\zeta_{\mathrm{L}}$ far from resonance but remains almost constant over a broad locking plateau near resonance ($d\zeta_0/d\zeta_{\mathrm{L}}\approx0$)~\cite{kondratiev2017SelfinjectionLockingLaser,voloshin2021DynamicsSolitonSelfinjection}. This locking behaviour can be understood through the CW solution branch of the system in the $(\zeta_{\mathrm{L}},\zeta_0)$ plane - also called the tuning or locking curve - whose stable branches are followed by the SIL system dynamics during a quasistatic detuning scan in the CW regime. When the locking plateau overlaps the soliton existence range in the effective-detuning parameter, self-injection locked soliton state is also accessible via a detuning scan~\cite{voloshin2021DynamicsSolitonSelfinjection}.

We perform detuning scan simulations of Eqs.~\ref{eq1:sil_lle} for an experimentally representative set of parameters, $d_2=0.01$, $\theta=0.6\pi$, $K=100$, $f^2=4$, and $\beta=0.01$. Using the transiently visited soliton states during the scan as initial guesses, we compute the underlying exact equilibrium soliton solutions by employing the recently developed dynamical systems methods and the associated Newton-Krylov algorithms~\cite{deshmukh2026NonlinearPeriodicOrbit}. Numerical continuation in the free-laser detuning parameter then yields the $N$-soliton solution branches (shown overlaid with the nonlinear tuning curve in Fig.~\ref{fig_1}(b)). The solution branches for different $N$ occupy distinct intervals of laser detuning $\zeta_{\mathrm{L}}$. Most consequentially, over a window $\zeta_{\mathrm{L}}\in[7.51,8.60]$ the single soliton branch is the only stable soliton solution branch. While this observation provides a possibility for deterministic single-soliton generation by laser tuning alone, in practice it might require a very precise control over the laser detuning.

Intracavity power profiles $|\Psi(\phi)|^2$ of the soliton solutions at a fixed free-laser detuning are equivalent to the single-LLE soliton profiles at the corresponding effective detuning, consistent with the LLE description for the forward field $\Psi(\phi, \tau)$ (Fig.~\ref{fig_1}(c), labeled I--VI (I--III unstable, IV--VI stable)). Consequently, a bifurcation diagram of the SIL soliton solution branches in the effective detuning parameter reveals the familiar foliated bifurcation structure of the LLE solitons (Fig.~\ref{fig_2}(a)). However, the stability properties of the SIL solitons are different compared to the corresponding single-LLE solitons owing to the possibility of instabilities induced by the coupling to the backward field $\rho_{\mathrm{B}}$ and the coupling to the laser cavity dynamics. We observe that in contrast to the $N$-soliton solution branches of the single-LLE that have stable segments over the entire detuning existence range for $f^2=4$~\cite{ParraRivas2014DynamicsLocalized}, the corresponding segments of the SIL $N$-soliton branches lose stability at lower effective detuning, and the maximum detuning $\zeta_{0,\mathrm{max}}^N$ for the stable $N$-soliton existence range of the SIL solitons decreases as $N$ increases. Since the soliton comb bandwidth increases with the effective detuning~\cite{herr2026FrequencyCombsCoherent}, the SIL single soliton attains a smaller maximum bandwidth than its single-LLE counterpart.

\begin{figure}[htbp]
\centering\includegraphics[width=0.8\columnwidth]{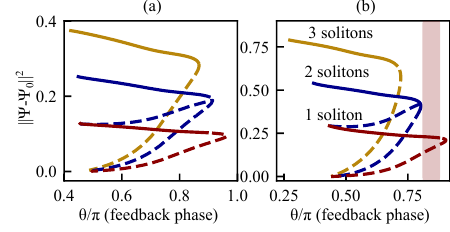}
\caption{\textbf{Slanted snaking in the feedback phase.} Bifurcation diagram of one-, two-, and three-soliton branches in the feedback phase parameter $\theta$ with (a) $d_2=0.01$ and (b) $d_2=0.05$ and fixed $\zeta_{\mathrm{L}}=3$, $K=100$, $f^2=4$, and $\beta=0.01$. Increasing $d_2$ increases the slant of the snaking structure and widens the feedback-phase interval in which the single-soliton branch remains stable after the higher-$N$ branches terminate; the shaded region in (b) marks the stable segment of the single-soliton-exclusive interval targeted in the direct numerical simulations.}
\label{fig_3}
\end{figure}

\textbf{Feedback phase study.}
The feedback phase $\theta$ controls the interference between the back-reflected field of the microresonator and the laser field and is a primary control parameter. We perform numerical continuation of the $N$-soliton solution branches in $\theta$ for a fixed value of free-laser detuning parameter $\zeta_L=3$ and the rest of the parameters fixed at $d_2=0.01$, $K=100$, $f^2=4$, and $\beta=0.01$. The bifurcation diagram in $\theta$ reveals a tilted ladder of the branches (Fig.~\ref{fig_3}(a))---often termed as \emph{slanted snaking}~\cite{Dawes2008SlantedSnaking}---in which the branches for different $N$ occupy distinct, offset ranges of $\theta$. Such slanting is a generic feature of snaking for pattern forming systems with non-local equations that support spatially localized patterns in a domain of finite size~\cite{Knobloch2016ConservationLaw,Firth2007HomoclinicSnaking}. In the SIL system the non-locality originates during the spatial averaging procedure of the forward intracavity field along the $\phi$ coordinate to obtain the mean field $\rho$ and power $P$ that enter the locking condition.

The soliton contribution in the cavity-averaged fields $\rho$ and $P$ is dictated by the dispersion parameter $d_2$. The slant is therefore expected to grow with $d_2$. The bifurcation diagram of the soliton branches computed at $d_2=0.05$ (Fig.~\ref{fig_3}(b)) confirms this, the single-soliton-exclusive interval in $\theta$ widens from $\sim0.05\pi$ at $d_2=0.01$ to $\sim0.1\pi$ at $d_2=0.05$.

To interpret this trend, we use an analytical approximation based on a $\mathrm{sech}$ ansatz of $N$ non-interacting solitons. In the $K\to \infty$ limit, the locking condition for this ansatz reduces to a relation between the effective detuning $\zeta_0$, the feedback phase $\theta$, and the soliton number $N$ (see Supplementary Information):
\begin{equation}
    \mathrm{Im}\!\left[
        \frac{e^{i\theta}\,\rho}
             {1 + i\zeta_0 - 2iP}
    \right] = 0,
    \label{eq:Ndks}
\end{equation}
where the cavity-averaged forward field $\rho$ and power $P$, evaluated for the $N$-soliton ansatz, depend on $\zeta_0$ and $N$. For each soliton number, this condition defines a phase curve $\theta_N(\zeta_0)$.

Comparing $N=1$ and $N=2$ curves gives the estimate $\Delta\theta_{\max}(d_2)=\max_{\zeta_0}\theta_1(\zeta_0)-\max_{\zeta_0}\theta_2(\zeta_0)$ for the single-soliton-exclusive phase range. Its increase with $d_2$ (Supplementary Information) suggests that microresonators with larger dispersion and Q-factor allow a broader range of feedback parameters for deterministic access to the single soliton.

\textbf{Dynamically accessing single soliton state.}
To demonstrate dynamical access to the single soliton states within the single-soliton-exclusive intervals predicted by the bifurcation analysis, we perform direct numerical simulations of Eqs.~\ref{eq1:sil_lle} under an adiabatic scan of the laser detuning and the feedback phase (other parameters fixed at $d_2=0.05, K=100, \; f^2=4,\; \beta=0.01$).

The protocol consists of two consecutive sweeps. Starting from a random initial condition, we first sweep the laser detuning linearly in time from $\zeta_{\mathrm{L}}=-47$ to $\zeta_{\mathrm{L}}=3$ at fixed feedback phase $\theta=0.6\pi$, followed by a sweep of the feedback phase from $\theta=0.6\pi$ to $0.85\pi$ with $\zeta_{\mathrm{L}}=3$ held constant. A small-amplitude noise is added periodically throughout the scan.

\begin{figure}[htbp]
\centering\includegraphics[width=0.8\columnwidth]{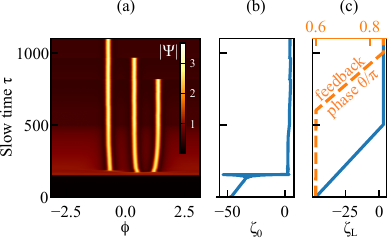}
\caption{\textbf{Dynamical access to a single soliton.} Direct numerical simulations of Eqs.~\ref{eq1:sil_lle} under a detuning scan from $\zeta_{\mathrm{L}}=-47$ to $3$ at fixed $\theta=0.6\pi$, followed by a feedback-phase scan to $\theta=0.85\pi$ at fixed $\zeta_{\mathrm{L}}=3$. (a) A spatiotemporal portrait of the forward field $|\Psi(\phi,\tau)|$ showing formation of a three-soliton state followed by sequential soliton loss, terminating in the single soliton state. (b) Corresponding effective detuning $\zeta_0(\tau)$, which initially tracks the free-laser detuning before entering the locking plateau near $\zeta_{\mathrm{L}}\simeq-31$ and remaining locked within the soliton existence range during the subsequent scan. (c) Applied scan trajectories for $\zeta_{\mathrm{L}}(\tau)$ and $\theta(\tau)/\pi$. Parameters are $d_2=0.05$, $K=100$, $f^2=4$, and $\beta=0.01$.}
\label{fig_4}
\end{figure}

During the first sweep, the forward field evolves from the CW state at $\zeta_{\mathrm{L}}=-47$ to a three-soliton state at the detuning $\zeta_{\mathrm{L}}=3$ (Fig.~\ref{fig_4}(a)), a detuning value where the three-soliton solution is stable. During the subsequent feedback-phase sweep, $\theta$ crosses the upper existence boundaries of the three- and two-soliton branches at $\theta=0.72\pi$ and $0.80\pi$, respectively (Fig.~\ref{fig_3}(b)). The system correspondingly loses solitons one by one at the boundary crossings until a single soliton remains at the terminal value of the feedback phase $\theta=0.85\pi$ which is within the single-soliton-exclusive range $\theta\in[0.80\pi, 0.90\pi]$. 

Repeated scans reproduce the same phenomenology, reliably terminating in a single-soliton state and thus establishing deterministic access to single solitons. These simulations confirm the dynamical relevance of the bifurcation structure: it governs the switching between $N$-soliton states and the selection of the stable single-soliton solution.

\textbf{Conclusion.}
We have analyzed the bifurcation structure of the soliton solutions of coupled LLE--laser equations describing a diode laser self-injection-locked to a microresonator in the weak-backscattering limit. Numerical continuation reveals a slanted snaking ladder of $N$-soliton branches in the free-laser detuning and feedback phase parameters. As a consequence, the feedback phase admits intervals in which only the single soliton solution exists. These intervals widen with the dispersion parameter $d_2$, a trend corroborated by an analytical approximation based on a $\mathrm{sech}$ $N$-soliton ansatz. Direct numerical simulations demonstrate dynamical access to the single soliton states using a detuning scan followed by a feedback-phase sweep.

We anticipate that these findings will inform the design principles of hybrid integrated soliton microcomb sources and provide guidance for deterministic single-soliton generation.

\begin{backmatter}

\bmsection{Acknowledgements}
The authors would like to thank Pooja Thakkar for her extensive help with designing the figures. 

\bmsection{Funding}
This work was supported by the Swiss National Science Foundation through SNSF Starting Grant (No. 234355) and the European Research Council (ERC) under the European Union’s Horizon 2020 research and innovation programme (grant no. 865677).
\bmsection{Disclosures}
The authors declare no conflicts of interest.

\bmsection{Data availability} Data underlying the results presented in this paper are not publicly available at this time but may be obtained from the authors upon reasonable request.

\end{backmatter}

\bibliography{sample}
\bibliographyfullrefs{sample}

\end{document}


\maketitle
\section*{Multiple Soliton Ansatz and Mean Field Derivation}

To estimate the single-soliton-exclusive interval in the strong feedback limit $K\to\infty$, we approximate the forward field by a CW background plus $N$ well separated solitons,
\begin{equation}
    \Psi(\phi)=\Psi_0+\sum_{j=1}^{N}S_j(\phi),
    \label{eq:ansatz_compact}
\end{equation}
with
\begin{equation}
    S_j(\phi)=
    \sqrt{2\zeta_0}\,
    \operatorname{sech}\!\left[
        \sqrt{\frac{2\zeta_0}{d_2}}(\phi-\phi_j)
    \right]e^{i\phi_s},
    \label{eq:single_soliton_ansatz}
\end{equation}
where the soliton phase is given by $\cos\phi_s=\frac{2\sqrt{2\zeta_0}}{\pi f}$ and the pulse centers obey the non-overlap condition
\begin{equation}
    |\phi_k-\phi_j|\gg
    \sqrt{\frac{d_2}{2\zeta_0}},
    \qquad j\neq k.
    \label{eq:non_interact}
\end{equation}
Thus overlap terms between different soliton peaks are neglected, and the finite-domain integrals over the soliton fields are replaced by their infinite-line values.

The background $\Psi_0$ is chosen on the lower homogeneous CW branch at the same effective detuning $\zeta_0$, given by the smallest positive solution of the cubic equation
\begin{equation}
    I^3-2\zeta_0I^2+(1+\zeta_0^2)I-f^2=0,
    \qquad I=|\Psi_0|^2.
    \label{eq:cw_cubic}
\end{equation}

The resulting mean field of this ansatz is
\begin{align}
    \rho
        &=\frac{1}{2\pi}\int_{-\pi}^{\pi}\Psi(\phi)\,d\phi
        \nonumber\\
        &\approx
        \Psi_0+\frac{N}{2}\sqrt{d_2}\,e^{i\phi_s}.
    \label{eq:rho}
\end{align}
For the power, we obtain
\begin{align}
    P
        &=\frac{1}{2\pi}\int_{-\pi}^{\pi}|\Psi|^2\,d\phi
        \nonumber\\
        &=
        |\Psi_0|^2
        +\frac{1}{2\pi}\sum_{j=1}^{N}
        \int_{-\pi}^{\pi}
        \left(
            \Psi_0^*S_j+\Psi_0S_j^*
        \right)d\phi
        \nonumber\\
        &\quad
        +\frac{1}{2\pi}
        \sum_{j,k=1}^{N}
        \int_{-\pi}^{\pi}S_jS_k^*\,d\phi.
    \label{eq:power_intermediate}
\end{align}
Neglecting the overlap terms $S_jS_k^*$ for $j\neq k$ and evaluating the single-soliton contributions gives
\begin{align}
    P
        &\approx
        |\Psi_0|^2
        +N\sqrt{d_2}\,
        \mathrm{Re}\!\left[\Psi_0e^{-i\phi_s}\right]
        +\frac{N\sqrt{2\zeta_0d_2}}{\pi}.
    \label{eq:power}
\end{align}

The backward field $\rho_{\mathrm{B}}$ is obtained from the steady-state backward-field equation in Eqs.~\ref{eq1:sil_lle},
\begin{equation}
    \rho_{\mathrm B}=
    \frac{i\beta\rho}{1+i\zeta_0-2iP}.
\end{equation}
Substituting this into the locking condition in Eqs.~\ref{eq1:sil_lle} yields
\begin{equation}
    \zeta_0-\zeta_{\mathrm L}
    =
    \frac{K}{f}
    \mathrm{Im}\!\left[
        \frac{e^{i\theta}\rho}{1+i(\zeta_0-2P)}
    \right].
    \label{eq:finite_K_locking}
\end{equation}
In the limit $K\to\infty$, the left-hand side of Eq.~\ref{eq:finite_K_locking} remains finite only if the imaginary part on the right-hand side vanishes:
\begin{equation}
    \mathrm{Im}\!\left[
        \frac{e^{i\theta}\rho}{1+i(\zeta_0-2P)}
    \right]=0.
    \label{eq:lock_cond}
\end{equation}

\begin{figure}[h]
    \centering
    \includegraphics[width=0.9\linewidth]{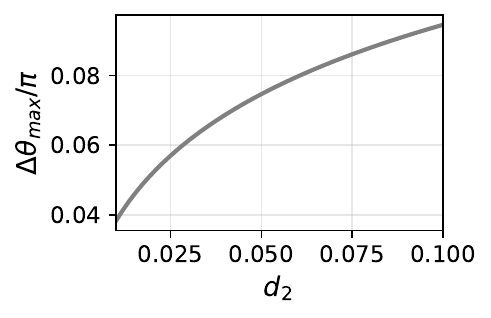}
    \caption{Analytical estimate of the single-soliton-exclusive feedback-phase interval. The normalized range $\Delta\theta_{\max}/\pi=[\max_{\zeta_0}\theta_1(\zeta_0)-\max_{\zeta_0}\theta_2(\zeta_0)]/\pi$ is estimated from the non-interacting $\mathrm{sech}$ ansatz as a function of the dispersion parameter $d_2$, for $f=2$ and the corresponding LLE soliton existence range $\zeta_0^{f=2}\in[2.72,\,4.93]$. The monotonic increase of the interval $\Delta\theta_{\max}$ in $d_2$ suggests that a larger dispersion parameter produces more favorable conditions for deterministic access to SIL single solitons.}
    \label{fig:delta_theta}
\end{figure}

We define
\begin{align}
    Q
        &=\zeta_0-2P
        \nonumber\\
        &=
        \zeta_0
        -2|\Psi_0|^2
        -2N\sqrt{d_2}\,
        \mathrm{Re}\!\left[\Psi_0e^{-i\phi_s}\right]
        -\frac{2N\sqrt{2\zeta_0d_2}}{\pi},
    \label{eq:Q_def}
    \\
    z_N(\zeta_0)
        &=
        \frac{
            \Psi_0+\dfrac{N}{2}\sqrt{d_2}\,e^{i\phi_s}
        }{1+iQ}.
    \label{eq:z_def}
\end{align}
Eq.~\ref{eq:lock_cond} is then
$\mathrm{Im}[e^{i\theta}z_N]=0$, so
\begin{equation}
    \theta_N(\zeta_0)=-\arg z_N(\zeta_0)
    \quad \mathrm{mod}\;\pi.
    \label{eq:theta_explicit}
\end{equation}
The single-soliton-exclusive phase interval is estimated as
\begin{equation}
    \Delta\theta_{\max}
    =
    \max_{\zeta_0}\theta_1(\zeta_0)
    -
    \max_{\zeta_0}\theta_2(\zeta_0).
    \label{eq:delta_theta}
\end{equation}
where the maximum is taken over the detuning values corresponding to the soliton existence range of the LLE for the given pump power $f^2$.
To study the dependence of this interval on the dispersion parameter, we evaluate Eq.~\ref{eq:delta_theta} as a function of $d_2$. Figure~\ref{fig:delta_theta} shows the resulting $\Delta\theta_{\max}(d_2)$ for $f=2$ and the corresponding $\zeta_0^{f=2}\in[2.72,\,4.93]$.

